\documentclass[a4paper]{article}
\usepackage{spconf,amsmath,graphicx}
\usepackage{amsmath,amssymb,mathtools,amsfonts}
\usepackage{graphicx,subfig,color,wrapfig,epsfig,epstopdf}
\usepackage{url}
\usepackage{multirow}
\usepackage{listings}
\usepackage{xspace}
\usepackage{cleveref}
\usepackage{INTERSPEECH2019}

\newcommand{\sync}{{SSGD}\xspace}

\newcommand{\dpsgd}{{DPSGD}\xspace}
\newcommand{\adpsgd}{{ADPSGD}\xspace}
\newcommand{\hadpsgd}{{H-ADPSGD}\xspace}
\newcommand{\swb}{{SWB}\xspace}
\newcommand{\ch}{{CH}\xspace}
\newcommand{\swba}{{SWB-300}\xspace}
\newcommand{\swbb}{{SWB-2000}\xspace}
\usepackage{INTERSPEECH2019}
\title{A Highly Efficient Distributed Deep Learning System For Automatic Speech Recognition}
\name{Wei Zhang, Xiaodong Cui, Ulrich Finkler, George Saon, Abdullah Kayi, Alper Buyuktosunoglu, Brian Kingsbury, David Kung, Michael Picheny}
\address{
  IBM Research
  }
\email{\{weiz,cuix,ufinkler,gsaon,kayi,alperb,bedk,kung,picheny\}@us.ibm.com}
\begin{document}

\maketitle
\begin{abstract}
 Modern Automatic Speech Recognition (ASR) systems rely on distributed deep learning to for quick training completion. To enable efficient distributed training, it is imperative that the training algorithms can converge with a large mini-batch size. 
 In this work, we discovered that Asynchronous Decentralized Parallel Stochastic Gradient Descent (ADPSGD) can work with much larger batch size than commonly used Synchronous SGD (SSGD) algorithm.  
 On commonly used public \swba and \swbb ASR datasets, ADPSGD can converge with a  batch size 3X as large as the one used in SSGD, thus enable training at a much larger scale. Further, we proposed a Hierarchical-ADPSGD (\hadpsgd) system in which learners on the same computing node construct a “super learner” via a fast allreduce implementation, and super learners deploy ADPSGD algorithm among themselves. On a 64 Nvidia V100 GPU cluster connected via a 100Gb/s Ethernet network, our system is able to train \swbb to reach a 7.6\% WER on the Hub5-2000 Switchboard (SWB) test-set and a 13.2\% WER on the Call-home (CH) test-set in 5.2 hours. To the best of our knowledge, this is the fastest ASR training system that attains this level of model accuracy for \swbb task to be ever reported in the literature. 

\end{abstract}

\section{Introduction}
Deep Learning (DL) drives the current  Automatic Speech Recognition (ASR) systems and has yielded models of unprecedented accuracy \cite{saon-interspeech-2017, msr-speech}. Stochastic Gradients Descent (SGD) and its variants are the de facto learning algorithms deployed in DL training systems. Distributed Deep Learning (DDL), which deploys different variants of parallel SGD algorithms in its core, is known today  as the most effective method to enable fast DL model training. Several DDL algorithms exist -- most notably, Synchronous SGD (SSGD) \cite{facebook-1hr}, parameter-server based Asynchronous SGD (ASGD) \cite{distbelief}, and Asynchronous Decentralized Parallel SGD (ADPSGD) \cite{adpsgd}. Among them, ASGD has lost its favor among practitioners due to its poor convergence behavior \cite{revisit-sync-sgd, facebook-1hr, deepspeech2, zhang2016icdm}. 
In our previous work \cite{icassp19}, we systematically studied the application of state-of-the-art  SSGD (\sync) and ADPSGD to challenging \swbb tasks. To date, it is commonly believed that \adpsgd and \sync converge with similar batch sizes \cite{adpsgd}. In this paper, we find \adpsgd can potentially smoothen objective function landscape and work with much larger batch sizes than \sync. As a result of this finding, we made the following contributions in this paper: (1) We systematically studied the convergence behavior of \sync and \adpsgd on two public ASR datasets -- \swba and \swbb -- with a state-of-the-art LSTM model and confirmed that \adpsgd enables distributed training on a much larger scale. To the best of our knowledge, this is the first demonstration conducted on large-scale public datasets that an asynchronous system can scale with a larger batch size than SSGD. (2) To reduce system staleness, improve system scalability w.r.t number of learners and  enable better communication and computation efficiency, we implemented a Hierarchical-ADPSGD (H-ADPSGD) system in which learners on the same computing node construct a ``super-learner'' via a fast allreduce\footnote{Allreduce is a broacast operation followed by a reduction operation (e.g., summation).} implementation and the super-learners communicate with each other in the \adpsgd fashion. Our system shortens the \swbb training from over 8 days on one V100 gpu to 5.2 hours on 64 gpus (i.e., 40X speedup), and the resulting model reaches 7.6\% WER on \swb and 13.2\% on \ch. To the best of our knowledge, this is the fastest system for state-of-the-art \swbb model training ever reported in the literature.

\section{Background and Problem Formulation}
\label{sec:back}
Consider the following stochastic optimization problem
\begin{align}
      \min_{\theta} F(\theta) =  \mathbb{E}_{\xi}[f(\theta;\xi)]
\end{align}
where $F$ is the objective function, $\theta$ is the parameters to be optimized (it is the weights of networks for DL) and $\xi\!\sim\!p(x)$ is a random variable on the training data $x$ obeying distribution $p(x)$. Supposing that there are $n$ training samples and $\xi$ assumes a uniform distribution,  $\xi\!\sim\!\text{Uniform}\{1, \cdots, n\}$,  we have
\begin{align}
      \min_{\theta} F(\theta) =  \mathbb{E}_{\xi}[f(\theta;\xi)] = \frac{1}{n}\sum_{i=1}^{n}f_{i}(\theta)
\end{align}
where $f_{i}(\theta)$ is evaluated at training sample $x_{i}$. In mini-batch based SGD, at each iteration $k$, we have
\begin{align}
      F(\theta_{k+1}) = F(\theta_{k}) - \alpha_{k}g(\theta_{k},\xi_{k})
\end{align}
where $\alpha_{k}$ is the learning rate and
\begin{align}
    g(\theta_{k},\xi_{k}) \triangleq \frac{1}{m}\sum_{s=1}^{m}\nabla f(\theta_{k},\xi_{k,s})
\end{align}
with $m$ being the size of the mini-batch.

\begin{figure}[t]
  \centering{
        \includegraphics[width=0.40\textwidth]{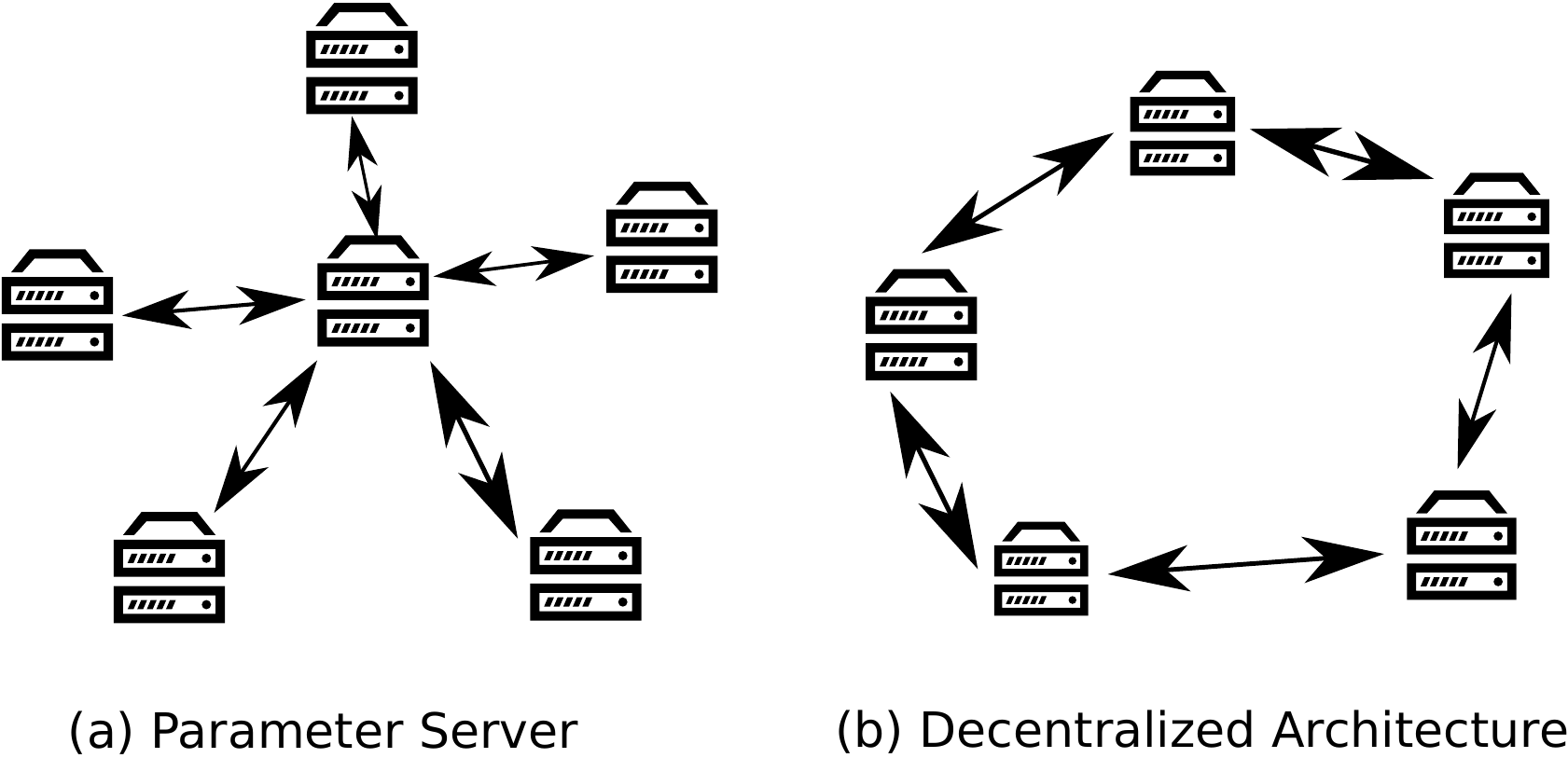}
  }
  \caption{A centralized distributed learning architecture (aka Parameter-Server architecture) and a decentralized distributed learning architecture}
  \label{fig:decen-graphs}
\end{figure}

DL problem is an optimization problem described above. DDL is the distributed computing paradigm that solves DL problems. At the dawn of DDL research, the Parameter Server (PS) architecture \cite{distbelief} was wildy popular. \Cref{fig:decen-graphs}(a) illustrates a PS design: each learner calculates gradients and sends them to the PS, before the PS sends the updated weights back to the learners.
The core of PS architecture is the ASGD algorithm that allows each learner to calculate gradients and asynchronously push/pull the gradients/weights to/from the PS. The weight update rule in ASGD is given in \Cref{eqn:update_async}:{
\begin{align}
    F(\theta_{k+1}) & = F(\theta_{k}) - \alpha_{k} g(\theta^{\zeta_{i}}_{k},\xi^{\zeta_{i}}_{k})  \label{eqn:update_async}
\end{align}
where $\zeta_{i} \in \{1, \cdots, \lambda\}$ is an indicator of the learner which made the model update at iteration $k$.
Soon researchers realized that such design led to sub-par models that can be trained fast but accuracy-wise significantly lag behind single-learner baseline because of the large system staleness issue \cite{zhang-ijcai-2016, revisit-sync-sgd, zhang2016icdm}.
SSGD regained its popularity mainly because the synchronization problems exacerbated by the cheap commodity systems could be addressed by decades of High Performance Computing (HPC) research. In SSGD, the weight update rule is given in \Cref{eqn:update_sync}
: each learner calculates gradients and receives updated weights in lockstep with the others.\footnote{Throughout this paper, we use $\lambda$ to denote the number of learners.} SSGD has exactly the same convergence rate as the single learner baseline while enjoying reasonable speedup. 

\begin{align}
    F(\theta_{k+1}) & = F(\theta_{k}) - \alpha_{k}\left[\frac{1}{\lambda}\sum_{i=1}^{\lambda} g(\theta^{i}_{k},\xi^{i}_{k})\right] \label{eqn:update_sync}
\end{align}

The summation and the following broadcast in \Cref{eqn:update_sync} is known as the ``All-Reduce'' operation in the HPC, which is a well-studied operation. When the message to be ``AllReduced'' is large, as in the DL case, an optimal algorithm exists that maximizes the communication bandwidth utilization \cite{fsu-allreduce}. Many incarnations of this algorithm exist, such as \cite{paddle-paddle,nccl,ddl}.
Like any synchronous algorithm, SSGD is subject to the \textit{straggler problem}, which means a slow learner slows down the entire training system.

Wildfire \cite{wildfire} is the first decentralized training system that has been demonstrated to work on modern deep learning tasks. The follow-up work in \cite{dpsgd, adpsgd} rigorously proved the decentralized parallel SGD algorithm in both synchronous and asynchronous forms can maintain the SSGD convergence rate while resolving the straggler problem. \Cref{fig:decen-graphs}(b) illustrates a decentralized parallel SGD system , where each learner $i$ calculates the gradients, updates its weights, and averages its weights with its neighbor $j$ in a ring topology. DPSGD/ADPSGD weights update rule is defined in \Cref{eqn:update_decentr}.

\begin{align}
    \Theta_{k+1} = \Theta_{k}T_{k} - \alpha_{k} g(\Theta_{k},\bm{\Xi}_{k})  \label{eqn:update_decentr}
\end{align}
where the columns of $\Theta_{k} = [\theta^{1}_{k}, \cdots, \theta^{\lambda}_{k}]$ are the weights of each learner; the columns of $\Xi_{k} = [\xi^{1}_{k}, \cdots, \xi^{\lambda}_{k}]$ are random variables associated with batch samples; the columns of $g(\Theta_{k},\bm{\Xi}_{k})  = [g(\theta^{1}_{k},\xi^{1}_{k}), \cdots, g(\theta^{\lambda}_{k},\xi^{\lambda}_{k})]$ are the gradients of each learner, and $T_{k}$ is a symmetric stochastic matrix (therefore doubly stochastic) with two $0.5$s at neighboring positions for each row and column and 0s everywhere else. This amounts to performing model averaging with one learner's left or right neighbor in the ring for each mini-batch weights update. The key difference between \dpsgd and \adpsgd is that \adpsgd allows overlapping of gradients calculation and weights averaging and removes the global barrier in \dpsgd that forces every learner to synchronize at the end of each minibatch training. Naturally, \adpsgd runs much faster than \dpsgd.

\section{Scalability of ADPSGD}
\label{sec:theory}


\begin{figure*}[t]
  \centering
  \subfloat[{Training loss comparison between \sync and \adpsgd, on SWB300 data-set}]
  {\includegraphics[width=2.0\columnwidth]{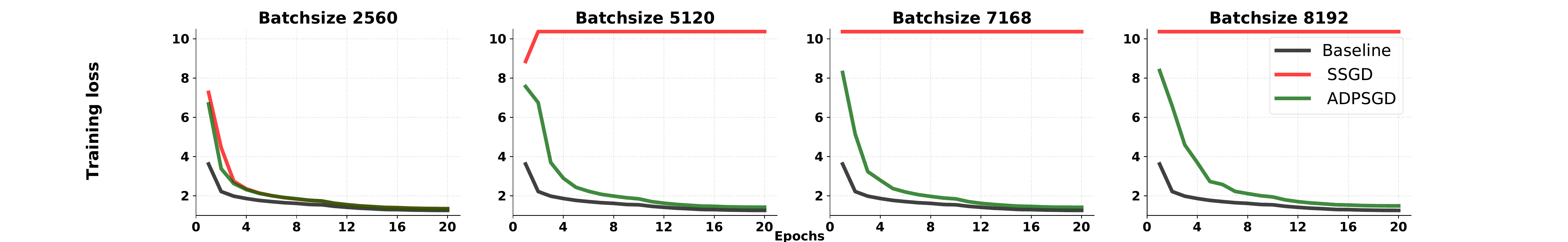}
    \label{fig:training-loss-swb300}
  }

  \subfloat[{Heldout loss comparison between \sync and \adpsgd, on SWB300 dataset}]
  {\includegraphics[width=2.0\columnwidth]{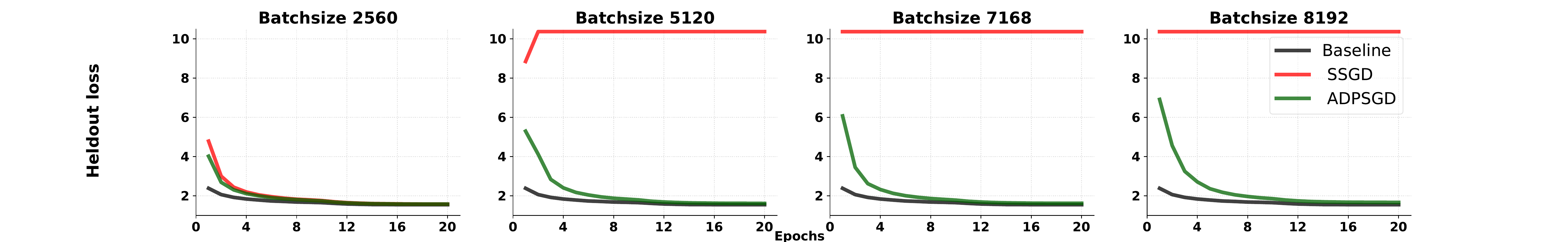}
    \label{fig:heldout-loss-swb300}
  }

  \subfloat[{Training loss comparison between \sync and \adpsgd on SWB2000 data-set}]
  {\includegraphics[width=2.0\columnwidth]{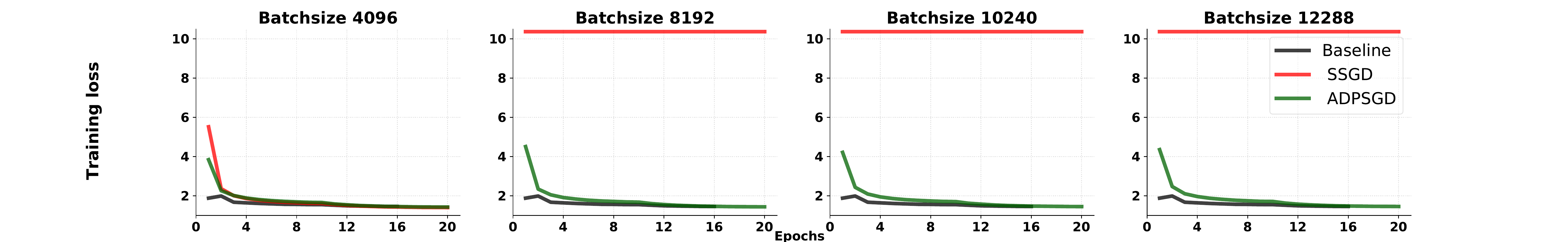}
    \label{fig:training-loss-swb2000}
  }

  \subfloat[{Heldout loss comparison between \sync and \adpsgd on SWB2000 data-set}]
  {\includegraphics[width=2.0\columnwidth]{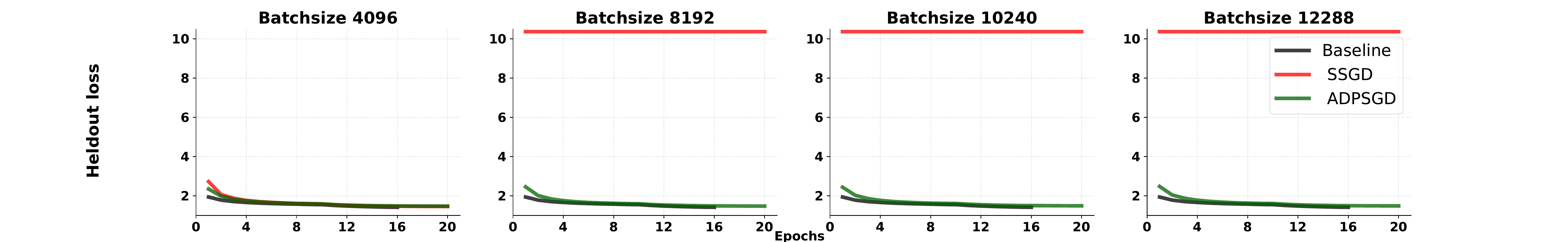}
    \label{fig:heldout-loss-swb2000}
  }
  \caption{Convergence comparison between \sync and \adpsgd on \swba and \swbb data-set. Baseline is plotted in black, \sync in red, \adpsgd in green. For \swba, \adpsgd converges similarly to the baseline up to batch size 8192, whereas \sync can only scale up to batch size 2560. Likewise, for \swbb, \adpsgd converges similarly to the baseline up to batch size 12288, whereas \sync can only scale up to batch size 4096. For all \sync and \adpsgd experiments, we use 16 learners.}
  \label{fig:conv}
\end{figure*}

It is well known that when batch-size is increased, it is difficult for a DDL system to maintain model accuracy \cite{facebook-1hr, zhang2016icdm}. In our previous work \cite{icassp19}, we designed a principled method to increase batch size while maintaining model accuracy with respect to training epochs for both \sync and \adpsgd on ASR tasks, up to batch size 2560. The key technique is as follows: (1) learning rate linear scaling by the ratio between effective batch size and baseline batch size. (2) linearly warmup the learning rate for the first 10 epochs before annealing at a rate $\frac{1}{\sqrt(2)}$ per epoch for the remaining epochs. For example, if the baseline batch size is 256 and learning rate is 0.1 and assume now the total batch size is 2560, then for the first 10 epochs, the learning rate increases by 0.1 every epoch before reaching 1.0 at the 10th epoch.
We demonstrated that \sync and \adpsgd can both scale up to the batch size of 2560 on ASR tasks in \cite{icassp19}.  In this paper, we adopt the same hyper-parameter setup and study the convergence behavior for \sync and \adpsgd under larger batch sizes.
\Cref{fig:training-loss-swb300} and \Cref{fig:heldout-loss-swb300} plot the training loss and held-out loss w.r.t epochs for \sync (in red) and \adpsgd (in green) under different batch sizes on \swba. The single-learner baseline with batch size 256 is plotted in black. \adpsgd can scale with a batch size 3x larger than \sync (i.e., 8192 vs 2560). \Cref{fig:training-loss-swb2000} and \Cref{fig:heldout-loss-swb2000} show a similar trend for \swbb -- \sync only scales up to batch size 4096 whereas \adpsgd scales up to batch size 12288. To make \sync converge with the same batch size as \adpsgd, one may use a much less aggressive learning rate (e.g. 4x smalle learning rate), which will take many more epochs to converge to the same level of accuracy.

\begin{table*}[t]
\centering
  \begin{tabular}{|l|l|l|l|l|l|l|l|l|l|}
\hline
\multirow{2}{*}{}    & BS256    & \multicolumn{2}{l|}{BS4096} & \multicolumn{2}{l|}{BS8192} & \multicolumn{2}{l|}{BS10240} & \multicolumn{2}{l|}{BS12288} \\ \cline{2-10}
    & Baseline & SSGD        & (H-)ADPSGD        & SSGD         & (H-)ADPSGD       & SSGD         & (H-)ADPSGD        & SSGD         & (H-)ADPSGD        \\ \hline
SWB & 7.5      & 7.6         & 7.5           & \textendash        & 7.6          & \textendash        & 7.5           & \textendash        & 7.8           \\ \hline
CH  & 13.0     & 13.1        & 13.2          & \textendash        & 13.2         & \textendash        & 13.5          & \textendash        & 13.6          \\ \hline
\end{tabular}
  \caption{WER comparison between baseline, \sync, and \adpsgd for different batch sizes. \textendash: not converged, BS: batch size.
    (H-)\adpsgd scales with 3X larger batch size than \sync. \adpsgd scales up to 16 GPUs, \hadpsgd scales up to 64 gpus.}
\label{tab:wer}
\end{table*}

We speculate \adpsgd can scale with a larger batch size than \sync for the following reason: ADPSGD is carried out with local gradient computation and update on each learner (second term on the RHS of \Cref{eqn:update_decentr}) and global model averaging (first term of the RHS of \Cref{eqn:update_decentr}) across the ring. The locally updated information is transferred to other learners in the ring by performing model averaging through neighboring learners. If we halt the local updates and only conduct the model averaging through the doubly stochastic matrix $T_{k}$, the system will converge to the equilibrium where all learners share the identical weights, which is $\frac{1}{n}\sum_{i=1}^{n}\theta^{i}_{k}$. After the convergence, each learner pulls the weights, computes the gradients, updates the model and conducts another round of model averaging to convergence. This amounts to synchronous SGD. Therefore, ADPSGD has synchronous SGD as its special case. However, in ADPSGD, local updates and model averaging run simultaneously. It is speculated that the local updates are carried out on $\frac{1}{n}\sum_{i=1}^{n}\theta^{i}_{k}$ perturbed by a noise term. This may give opportunities to use a larger batch size that is not possible for synchronous SGD.

\section{Design of Hierarchical ADPSGD}
\label{sec:design}
\begin{figure}
  \centering
    {\includegraphics[width=1.0\columnwidth]{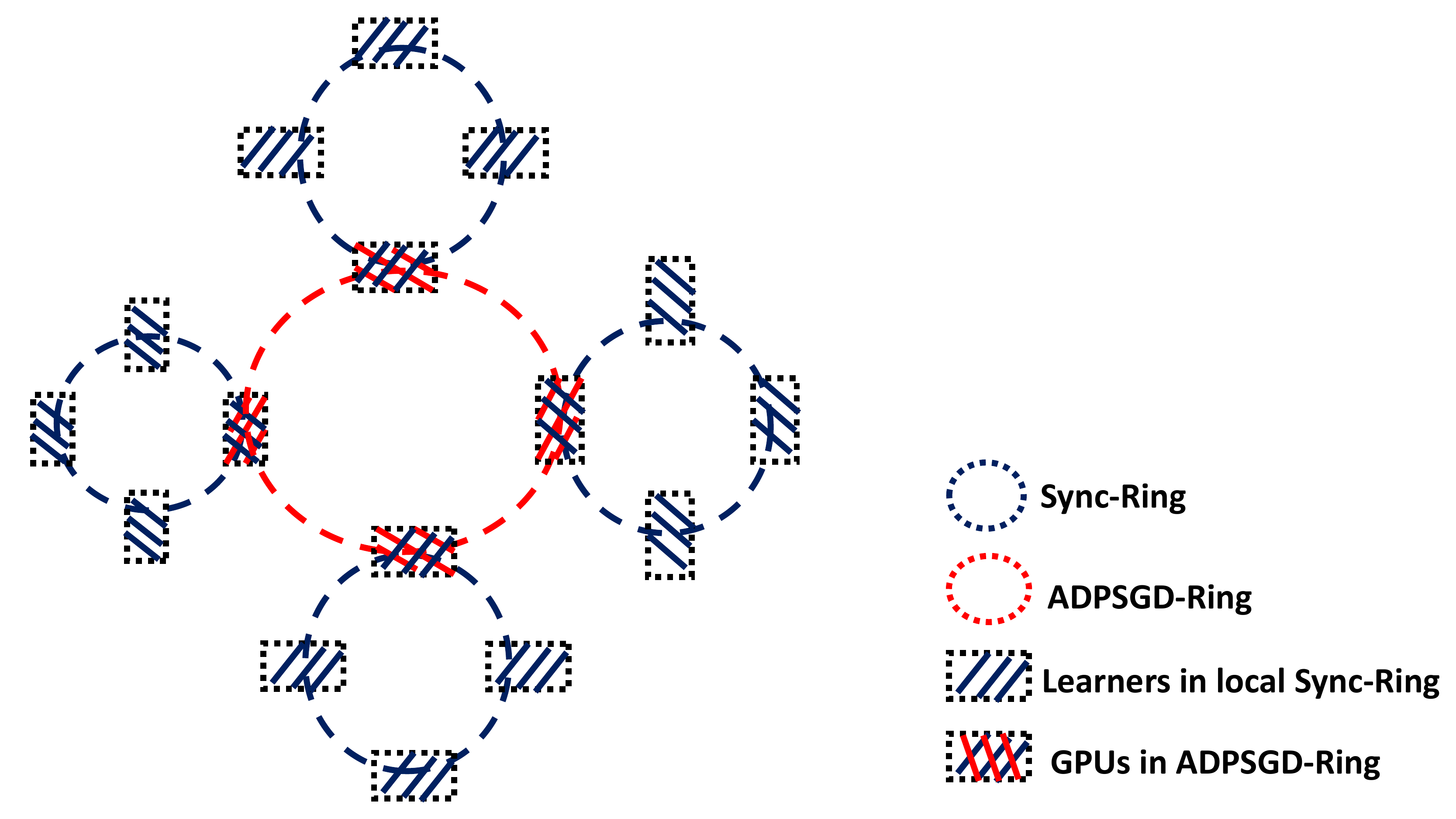}
      \caption{System architecture for \hadpsgd. Learners, on the same computing node, constitute a super-learner via a synchronous ring (blue). The super-learners constructs the \adpsgd ring (red).}
    \label{fig:arch}
    }
\end{figure}
In practice, \adpsgd sees signficant accuracy drop when scaling over more than 16 learners on ASR tasks \cite{adpsgd, icassp19} due to system staleness issue. We built \hadpsgd, which is a hierarchical system as depicted in \Cref{fig:arch}, to address the staleness issue. $N$ learners constructs a super-learner, which applies the weight update rule as in \Cref{eqn:update_sync}. The staleness across learners in this super-learner is effectively 0. 
Each super-learner then participates in the \adpsgd ring and update its weights using \Cref{eqn:update_decentr}. Further, assume there are a total of $\lambda$ learners; then \textit{the convergence behavior of this $\lambda$-learner \hadpsgd system is equivalent to that of a $\frac{\lambda}{N}$-learner \adpsgd with the same total batch size}. This design ensures \hadpsgd scales with $N$ more learners than its \adpsgd counterpart, while maintaining the same level of model accuracy. Moreover, \hadpsgd improves communication efficiency and  
reduces main memory traffic and CPU pressure across super-learners.

\section{Methodology}
\label{sec:meth}

\subsection{Software and Hardware}
PyTorch 0.4.1 is our DL framework. Our communication library is built with CUDA 9.2 compiler, the CUDA-aware OpenMPI 3.1.1, and g++ 4.8.5 compiler. We run our experiments on a 64-GPU 8-server cluster. Each server has 2 sockets and 9 cores per socket. Each core is an Intel Xeon E5-2697 2.3GHz processor. Each server is equipped with 1TB main memory and 8 V100 GPUs. Between servers are 100Gbit/s Ethernet connections. GPUs and CPUs are connected via PCIe Gen3 bus, which has a 16GB/s peak bandwidth in each direction per socket. 




\subsection{DL Models and Dataset}
\label{sec:meth:model}
The acoustic model is an LSTM with 6 bi-directional layers. Each layer contains 1,024 cells (512 cells in each direction). On top of the LSTM layers, there is a linear projection layer with 256 hidden units, followed by a softmax output layer with 32,000 units corresponding to context-dependent HMM states. The LSTM is unrolled with 21 frames and trained with non-overlapping feature subsequences of that length.  The feature input is a fusion of FMLLR (40-dim), i-Vector (100-dim), and logmel with its delta and double delta (40-dim $\times$3). This model contains over 43 million parameters and is about 165MB large.

The language model (LM) is built using publicly available training data, including Switchboard, Fisher, Gigaword, and Broadcast News, and Conversations. Its vocabulary has 85K words, and it has 36M 4-grams. 
The two training datasets used are \swba, which contains over 300 hours of training data and has a capacity of 30GB, and \swbb, which contains over 2000 hours of training data and has a capacity of 216GB. The two training data-sets are stored in HDF5 data format. The test set is the Hub5 2000 evaluation set, composed of two parts: 2.1 hours of switchboard (SWB) data and 1.6 hours of call-home (CH) data.

\section{Experimental Results}
\label{sec:results}
\subsection{Convergence Results}
\Cref{tab:wer} records the WER of \swbb models trained by \sync and \adpsgd under different batch sizes. Single-gpu training baseline is also given as a reference. \textit{\adpsgd can converge with a batch size 3x larger than that of \sync, while maintaining model accuracy. }

\subsection{Speedup}
\begin{figure}
  \centering
      {\includegraphics[width=0.6\columnwidth]{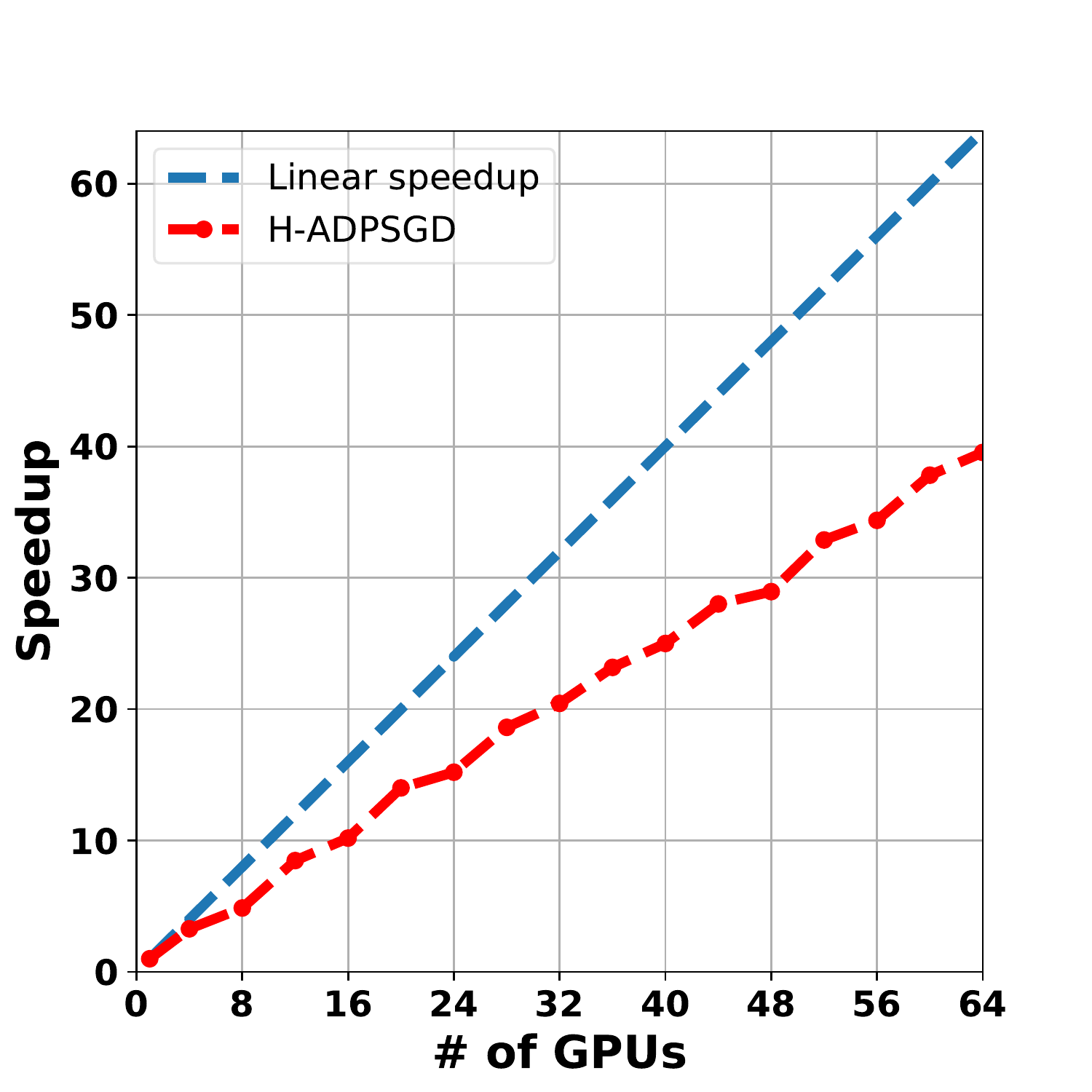}
        \caption{Speedup up to 64 gpus. Batch-size per gpu is 128.}
    \label{fig:speedup-x86}
    }
\end{figure}

\Cref{fig:speedup-x86} shows the \hadpsgd speedup. Using a batch size 128 per gpu, our system achieves 40X speedup over 64 gpus. It takes 203 hours to finish \swbb training in 16 epochs on one V100 GPU. \textit{It takes \hadpsgd 5.2 hours to train for 16 epochs to reach WER 7.6\% on SWB and WER 13.2\% on CH, on 64 gpus. }



\section{Conclusion and Future Work}
\label{sec:future}
In this work, we made the following contributions: (1) We discovered that \adpsgd can scale with much larger batch sizes than the commonly used SSGD algorithm for ASR tasks. To the best of our knowledge, this is the first asynchronous system that scales with larger batch sizes than a synchronous system for public large-scale DL tasks. (2) To make \adpsgd more scalable w.r.t number of learners, we designed a new algorithm \hadpsgd, which trains a \swbb model to reach WER 7.6\% on SWB and WER 13.2\% on CH in 5.2 hours, on 64 Nvidia V100 GPUs. To the best of our knowledge, this is the fastest system that trains \swbb to this level of accuracy. We plan to investigating the theoretical justification for why \adpsgd scales with a larger batch size than \sync in our future work.

\section{Related Work}
\label{sec:related}
DDL systems enable many AI applications with unprecedented accuracy, such as speech recognition \cite{deepspeech2,bmuf}, computer vision \cite{facebook-1hr}, language modeling \cite{nvidia-lm-scaling}, and machine translation \cite{fb-mt-scaling}. During the early days of DDL system research, researchers could only rely on loosely-coupled inexpensive computing systems and adopted PS-based ASGD algorithm \cite{distbelief}. ASGD has inferior model performance when learners are many, and SSGD has now become a mainstream DL training method \cite{revisit-sync-sgd, facebook-1hr, deepspeech2, zhang2016icdm, gadei}. Consequently, these systems encounter the classical straggler problem in distributed system design. Recently, researchers proposed ADPSGD, which is proved to have the same convergence rate as SSGD while completely eliminating the straggler problem \cite{adpsgd}.  Previous work \cite{icassp19} applied \adpsgd to the challenging SWB2000-LSTM task to achieve state-of-the-art model accuracy in a record time (11.5 hours). In this work, we discovered that \adpsgd can scale with much larger batch sizes than \sync and designed a hierarchical ADPSGD training system that further improves training efficiency, with the resulting system improving training speed over \cite{icassp19} by over 2X.

\bibliographystyle{IEEEtran}

\bibliography{mybib}

\end{document}